\documentclass[twocolumn,showpacs,aps,prb,superscriptaddress,floatfix]{revtex4}

\usepackage{graphicx}
\usepackage{dcolumn}
\usepackage{bm}

\begin{document}

\preprint{}

\newcommand{\scbo}{SrCu$_2$(BO$_3$)$_2$ }
\newcommand{\Imchi}{Im $\chi$({\bf Q}, $\hbar \omega$) } 

\affiliation{Department of Physics and Astronomy, McMaster University,
Hamilton, Ontario, L8S 4M1, Canada}
\affiliation{Canadian Institute for Advanced Research, 180 Dundas St. W., 
Toronto, Ontario, M5G 1Z8, Canada} 

\author{B.D. Gaulin}
\affiliation{Department of Physics and Astronomy, McMaster University,
Hamilton, Ontario, L8S 4M1, Canada}
\affiliation{Canadian Institute for Advanced Research, 180 Dundas St. W.,
Toronto, Ontario, M5G 1Z8, Canada}
\author{S.H. Lee}
\affiliation{National Institute of Standards and Technology, 100 Bureau 
Dr., Gaithersburg, MD, 20899-8562, U.S.A.} 
\author{S. Haravifard}
\affiliation{Department of Physics and Astronomy, McMaster University,
Hamilton, Ontario, L8S 4M1, Canada}
\author{J.P. Castellan}
\affiliation{Department of Physics and Astronomy, McMaster University,
Hamilton, Ontario, L8S 4M1, Canada}
\author{A.J. Berlinsky}
\affiliation{Department of Physics and Astronomy, McMaster University,
Hamilton, Ontario, L8S 4M1, Canada}
\affiliation{Canadian Institute for Advanced Research, 180 Dundas St. W., 
Toronto, Ontario, M5G 1Z8, Canada} 
\author{H.A. Dabkowska}
\affiliation{Department of Physics and Astronomy, McMaster University,
Hamilton, Ontario, L8S 4M1, Canada}
\author{Y. Qiu}
\affiliation{National Institute of Standards and Technology, 100 Bureau 
Dr., Gaithersburg, MD, 20899-8562, U.S.A.} 
\affiliation{Department of Materials Science and Engineering, University
of Maryland, College Park, MD, 20742, U.S.A.}
\author{J.R.D. Copley}
\affiliation{National Institute of Standards and Technology, 100 Bureau
Dr., Gaithersburg, MD, 20899-8562, U.S.A.}%

\title{High Resolution Study of Spin Excitations\\ in the 
Shastry-Sutherland Singlet Ground State of \scbo} 

\begin{abstract} 
High resolution, inelastic neutron
scattering measurements on \scbo reveal the dispersion of the three single
triplet excitations continuously across the (H,0) direction within
its tetragonal basal plane.  These measurements also show distinct {\bf Q}
dependencies for the single and multiple triplet excitations, and
that these excitations are largely dispersionless perpendicular to this
plane.  The temperature dependence of the intensities of these excitations
is well described as the complement of the dc-susceptibility of \scbo.
\end{abstract} \pacs{75.10.Jm, 75.25.+z, 75.40.Gb}

\maketitle 

Quantum magnets which display collective singlet ground states have been
of much recent interest \cite{reviews}.  Several of these, such as
CuGeO$_3$ \cite{CuGeO}, MEM-(TCNQ)$_2$ \cite{MEM}, and NaV$_2$O$_5$
\cite{NaVO} result from spin-Peierls phenomena in low dimensions where the
lattice combines with s=1/2 spin degrees of freedom to break translational
symmetry below some characteristic phase transition temperature, and a
non-magnetic ground state with a characteristic energy gap is observed.  A
related state is observed in the s=1 antiferromagnetic chain compounds,
such as CsNiCl$_3$ \cite{CsNiCl} and NENP \cite{NENP}, where a
non-magnetic ground state with an energy gap, the Haldane gap, forms in
the absence of translational symmetry breaking.

\scbo has been proposed \cite{Ueda, Kageyama2} as a realization of the
Shastry-Sutherland model \cite{Shastry} of interacting dimers in two
dimensions.  This material crystallizes \cite{Structure} into the 
tetragonal space group I$\bar{4}$2m  with lattice parameters a=8.995 
$\AA$, c=6.649 $\AA$.  
Magnetically, it can be thought of in terms of well isolated basal planes
populated by antiferromagnetically coupled s=1/2 moments on the Cu$^{2+}$
sites.  These are arranged in dimers at right angles to each other and
forming a square lattice.  The microscopic Hamiltonian appropriate to this
system is based on:  \begin{eqnarray} \mathcal{H} & = & J\sum_{nn}{\bf
S_{i}\cdot S_{j}} + J'\sum_{nnn}{\bf{S_{i}\cdot S_{j}}} \end{eqnarray}

It is known that both interactions, J and J', are antiferromagnetic and
similar in strength, such that the system is not far removed from the
critical value of x=J'/J known to be appropriate to the quantum phase
transition which separates a 4 sublattice Neel state from a collective
singlet ground state.  

The estimated values of J and J' have evolved over time as both theory and 
experiment have improved.  Early estimates of J and x were J=8.6 meV 
and x=0.68 \cite{Ueda}, very close to the critical value x$_c$=0.69 
\cite{Weihong}, where the single triplet excitation goes soft.
More recently Miyahara and Ueda \cite{Ueda2} found J=7.3 meV and 
x=0.635 from fits to the magnetic susceptibility.  Somewhat smaller values 
were obtained by Knetter et al \cite{Knetter1} who compared theoretical 
and experimental 
ratios of the energy of the lowest S=1 two triplet bound state 
to the single triplet gap, to obtain J=6.16 meV and 
x=0.603.  Note that within this theory \cite{Knetter1}, the 
collective singlet ground state becomes unstable when the lowest energy 
{\it two triplet bound state} goes soft.

Earlier, relatively low resolution inelastic neutron scattering
measurements \cite{Kageyama, Cepas, Kakurai}have directly identified three 
bands of
excitations corresponding to single (n=1) triplet excitations, as well as
to two (n=2), and to three (n=3) triplet excitations. These measurements
directly show the appearance of the energy gap in the spectrum of
excitations with decreasing temperature, as well as the dispersion of
these excitations in an applied magnetic field.

The more recent of these studies \cite{Cepas, Kakurai} has been able to
investigate subleading terms in the spin Hamiltonian.  Terms such as
the Dzyaloshinski-Moriya (DM) interaction, which is allowed by symmetry
between spins on neighboring dimers, and anisotropic exchange, can be
responsible for both dispersion of these excitations, and the removal of
the three-fold degeneracy which would otherwise characterize the
n=1 triplet excitation spectrum in \scbo.  The presence of such small
terms in the spin Hamiltonian has also been investigated through high
field specific heat measurements \cite{Jorge}, performed on samples from
the same crystal growth as that used in the present study, and through ESR
measurements \cite {Zorko, Nojiri}.

In this letter, we report new high resolution inelastic neutron scattering
measurements, which probe both the energy and {\bf Q} dependencies of
these previously identified bands of excitations with new precision.  
Measurements were performed on two different high resolution cold neutron
instruments, allowing both sufficiently high energy resolution to resolve
the three n=1 triplet excitations in \scbo, and sufficiently
high {\bf Q} resolution to discern different {\bf Q}-dependencies of the
n-triplet excitations, where n=1,2,3.

The present single crystal of SrCu$_2$($^{11}$BO$_3$)$_2$ was grown from a
self-flux by floating zone image furnace techniques in an O$_2$
atomosphere using a four mirror furnace.  It is cylindrical in shape,
with approximate dimensions 0.6 cm in diameter by 10 cm long, and an
a-axis was roughly aligned along the cylindrical axis.  Small pieces of
the crystal were used for bulk characterization, and the characteristic
fall off of the dc susceptibility near 10 K, as well as well defined steps
in the magnetization versus applied magnetic field at high field, were
observed \cite{Jorge}.  Neutron diffraction measurements, enabled by the
use of $^{11}$B isotope, revealed a high quality single crystal throughout
the volume of the sample with a mosaic spread of less than 0.2 degrees.

The crystal was mounted in a pumped $^4$He cryostat with the long
cylindrical axis vertical, placing the (H,0,L) plane of the crystal
coincident with the horizontal plane, and hence the scattering plane.  
Neutron scattering measurements were performed using the Disk Chopper
Spectrometer (DCS) and the SPINS triple axis spectrometer, both located on
cold neutron guides at the NIST Center for Neutron Research.

The DCS uses choppers to create pulses of monochromatic neutrons whose
energy transfers on scattering are determined from their arrival times
in the instrument's 913 detectors located at scattering angles from -30
to 140 degrees. Using 5.1 meV incident neutrons and the medium
resolution option \cite{Copley}, the energy resolution is ~0.09 meV.
\begin{figure}[t] \centering
\includegraphics[height=4in]{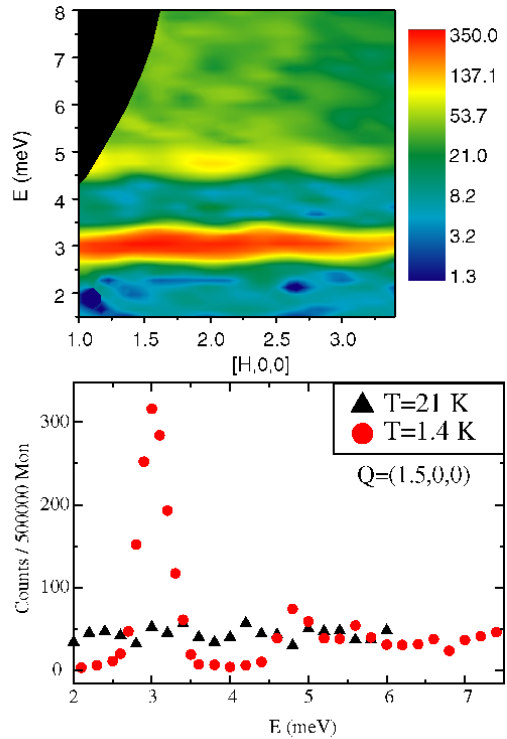} 
\caption{Top panel: A map of the measured dynamic structure factor for 
\scbo at T=1.4 K along the (H,0,0) direction is shown.  The intensity 
scale is logarithimic.  The map is made 
up of constant-{\bf Q} scans of the form shown in the bottom panel, where 
{\bf Q}=(1.5,0,0).} 
\label{fig1} 
\end{figure}

The SPINS triple axis spectrometer was operated using seven pyrolitic
graphite analyser blades accepting 7 degrees in scattering angle, and
neutrons of 5 meV, fixed scattered energy.  A cooled Be filter was
placed in the scattered beam to remove contamination of higher order
neutrons, and the resulting energy resolution was $\sim$ 0.5 meV.

\begin{figure}[tbp]
\centering
\includegraphics[width=.8\columnwidth]{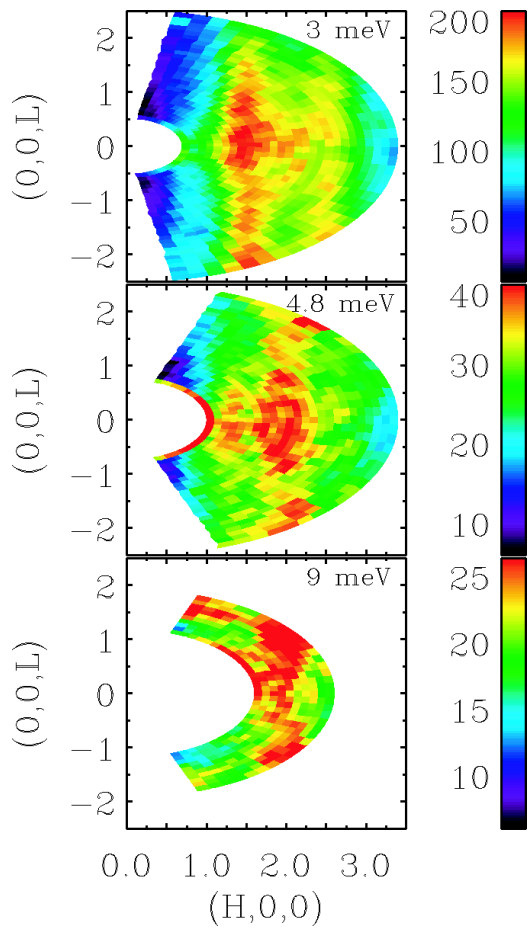} 
\caption{Constant energy scans at 3 (top panel), 4.8 meV (middle panel) 
and 9 meV (bottom panel) probing the {\bf Q}-dependence of the n=1,2, and 
3 triplet excitations in \scbo within the (H,0,L) plane at T=1.4 
K.} 
\label{fig2} 
\end{figure}
A program of constant-{\bf Q} measurements across the (H,0,L) plane of
\scbo at 1.4 K was carried out covering energies from 1 to 8 meV.  A
compendium of such scans with intervals of $\Delta$H=0.2 is shown in the
color contour map in Fig. 1a, which displays data within the $\hbar
\omega$, H plane at L=0.  Repesentative scans making up this map are shown
in Fig. 1b.  One
clearly identifies both the n=1 triplet excitation near $\hbar
\omega$=3.0 meV and the n=2 triplet excitation, whose spectral
weight is maximum near $\hbar \omega$$\sim$ 4.9 meV.  One also sees a
continuous component to the n=2 triplet excitation, which extends
to the highest energies collected in this set of measurments, 8 meV.

\begin{figure}[t] \centering
\includegraphics[width=1.0\columnwidth]{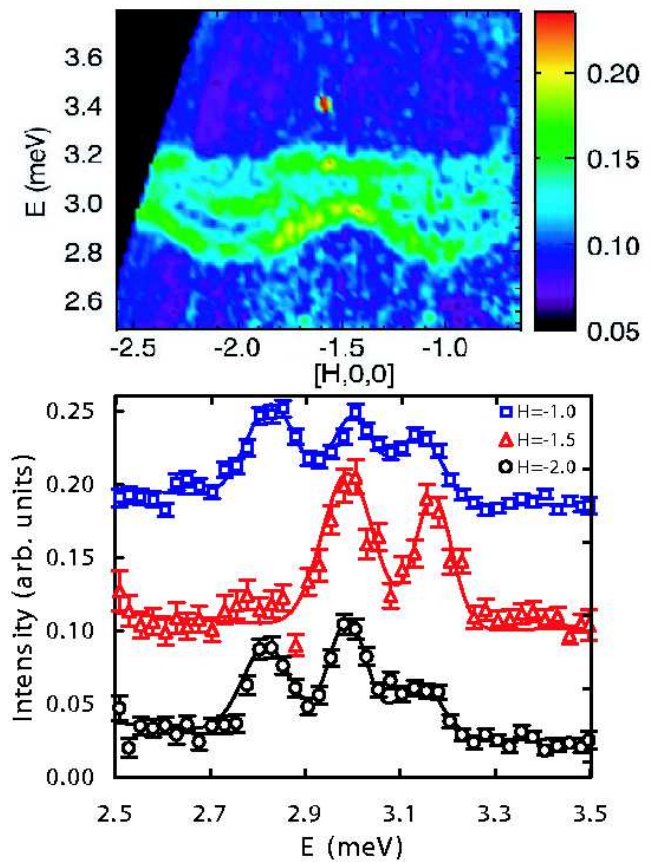} 
\caption{Top panel: Colour contour map of the dynamic structure factor 
for the n=1 triplet excitations along the (H,0) direction within 
the basal plane of \scbo.  These measurements, taken with the DCS 
spectrometer, integrate along L, as described in the text.  Bottom panel: 
Cuts through the map of the top panel are shown, which approximate 
constant-{bf Q} scans and clearly resolve the three 
branches to the n=1 triplet excitation.  The solid lines are Gaussian fits 
to the data.} \label{fig3} \end{figure}

These measurements are qualitatively similar to those first reported by
Kageyama et al. \cite{Kageyama}, within the (H,K,0) plane of \scbo and
with lower energy resolution; however there are important differences.  
For example, as can be seen in Fig. 1, we observe no substantial
dispersion of the maximum of the spectral weight of the n=2
triplet excitation in contrast to a bandwidth of 1.5 meV reported
in these earlier measurements \cite{Kageyama}.  Also, as we discuss next,
the {\bf Q}-dependence of these excitations is different from that
reported by Kageyama et al \cite{Kageyama}.

\begin{figure}[t] \centering
\includegraphics[width=1.0\columnwidth]{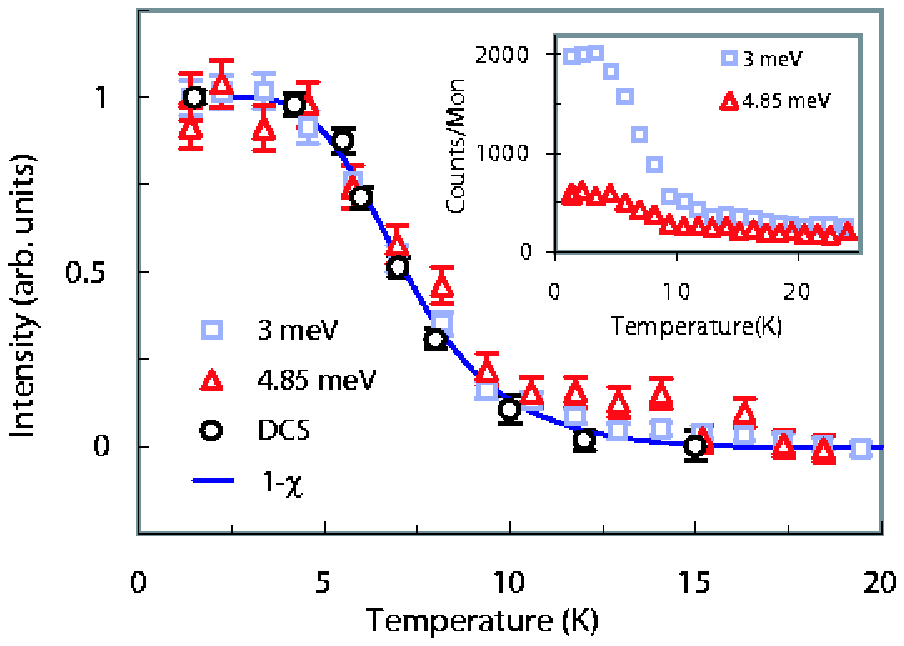} 
\caption{The temperature dependence of inelastic intensity at (1.5,0,0) 
and $\hbar \omega$ =3 meV, as well as at (2,0,0) and $\hbar \omega$=4.8 meV are 
shown.  The inset shows the raw intensity data, while the main figure 
shows the normalized intensity assuming zero at 25 K.  This is compared to 
the complement of the measured dc susceptibility (1-$\chi$), which clearly 
describes well the observed temperature dependence.} \label{fig4} 
\end{figure}

Figure 2 shows the results of constant energy scans performed at
$\hbar\omega$=3, 4.8 and 9 meV, corresponding to the n-triplet
excitations with n=1,2,3 respectively.  These measurements clearly show
the n=1 excitation at 3 meV to peak up at half integer values of H, that
is at H=1.5 and 2.5, in maps of this scattering within the (H,0,L) plane,
and at integer values of H=2 for the n=2 excitation at 4.8 meV and for the
n=3 excitation at 9 meV. These results indicate distinct form factors for
the n-triplet excitations, with the n=1
triplet being different from the multi-triplet
excitations.  Knetter and Uhrig \cite{Knetter} have recently calculated
the n=2 triplet contribution to the dynamic structure factor
within the (H,K,0) plane of \scbo by perturbative techniques.  They show 
that it is peaked at (2,0,0), consistent with the present set of
inelastic scattering measurements.

It is also clear from Fig. 2 that the {\bf Q}-dependence of all the
excitations show little L dependence, consistent with well isolated two
dimensional basal planes. Given that the scattering displays very little
L-dependence, the DCS measurements can be integrated along L, resulting in
a good quality determination of the dispersion of the n=1 triplet
excitations in the (H,0) direction within the tetragonal basal plane.  
This is what is shown in Fig. 3, where the top panel shows a color contour
map of the inelastic scattering acquired using 5.1 meV incident neutrons.  
The bottom three panels show cuts through this map, which approximate
constant-{\bf Q} scans at (-1,0), (-1.5,0) and (-2,0), from top to bottom,
respectively.

These inelastic measurements clearly resolve three branches of
excitations, corresponding to S$^z$=0,$\pm$1 transitions from the ground
state, the lower two of which are nearly degenerate at (-1.5,0).  These
measurements have the advantage that they track the dispersion of the
single singlet-triplet excitations {\it continuously} in {\bf Q} along
(H,0).  This color contour map is also consistent with the mapping of the
Q-dependence of the integral of the singlet-triplet scattering, shown in
Fig. 1a, which shows this scattering to peak in intensity near (-1.5,0).

Although the high resolution n=1 triplet dispersion curves
shown in Fig. 3 have not been previously reported, a substantial amount of
information is known about these spectra from ESR measurements.  
In particular Nojiri et al. \cite{Nojiri} used
ESR to measure the q=0 n=1 triplet excitations finding states at
2.81 meV (679 $\pm$ 2 GHz) and 3.16 meV (764 $\pm$ 2 GHz), in excellent
agreement with the lowest and highest n=1 triplet excitations at
{$\bf Q$}=(-1,0) and (-2,0) shown in Fig. 3.  The splitting between these
two excitations has been shown to be 4D \cite{Cepas}, where D is the
inter-dimer DM interaction.  In fact the dispersion
relation shown in Fig. 3 bears similarities to the calculation shown in 
Fig. 2c
of Cepas et al. \cite{Cepas}, except for the small gap, $\sim$ 0.2 meV, we
observe between the S$^z$=0 transition and the upper of the S$^z$=$\pm$1
branches at {$\bf Q$}=(-1.5,0).  This gap must result from some other
anisotropic interaction, such as the recently proposed \cite {Zorko}
perpendicular intra-dimer DM interaction. 

We note that the band width of the middle, S$^z$=0 mode in Fig. 3 is 
extremely small, roughly 0.05 meV.  This mode is the least affected by 
anisotropic interactions, and hence it is the mode most directly 
comparable to calculations of the n=1 triplet excitation based 
on Eq(1).  This bandwidth, which scales like x$^6$, is similar to the 
bandwidth found by Weihong et al. \cite{Weihong} for x=0.6, which is about 
4$\%$ of the n=1 triplet gap energy.

We have also measured the temperature dependence of both the n=1 and n=2
triplet excitations with the SPINS triple axis spectrometer, and that of 
the n=1 triplet excitations with DCS. 
The SPINS measurements at {\bf Q}=(1.5,0,0) and (2,0,0)  and energy
transfers of $\hbar \omega$ =3 meV and 4.9 meV, sample the
n=1 and n=2 triplet excitatons, respectively, and are shown in Fig. 4.
The DCS measurements integrate the inelastic scattering in data 
sets of the form shown in the top panel of Fig. 3.  The DCS 
measurements integrate between 2.7 and 3.3 meV and across all wavevectors 
from H=-2.25 to H=-0.75 along (H,0) within the basal plane. 

Figure 4 shows that the temperature dependence of the n=1 and n=2
triplet excitations are identical, which may not have been
concluded from earlier measurements \cite{Kageyama}. We do however confirm
the very rapid dropoff of this inelastic intensity with increasing
temperature.  Such a strong temperature dependence is not characteristic
of a system undergoing a phase transition, and indeed there is no evidence
of a broken symmetry associated with the appearance of the collective
singlet ground state. Rather the temperature dependence of the inelastic
scattering can be very well described as the complement of the dc
susceptibility.  That is that the temperature dependence to the inelastic
scattering follows that of the dc susceptibility.  This has been measured
and modelled using finite temperature Lanczos method \cite{Jorge}.  The
complement of $\chi$, referred to as 1-$\chi$ in Fig. 4, is 
given by $\chi$(T=20 K)-$\chi$(T).  This quantity is scaled to compare to
the temperature dependence of the inelastic scattering, and one can see
that this provides an excellent description of the temperature dependence
of the inelastic scattering.

We wish to acknowledge the contribution of G. Gu to this study.  This 
work was supported by NSERC of Canada, and utilized facilities supported 
in part by the National Science Foundation under Agreements DMR-9986442 
and DMR-0086210.

\end{document}